\documentclass[12pt]{article}\usepackage{amsmath,amsfonts,epsfig}
\newcommand{\bse}{\begin{subequations}}
\newcommand{\ese}{\end{subequations}}
\newcommand{\be}{\begin{equation}}
\newcommand{\ee}{\end{equation}}
\newcommand{\bea}{\begin{eqnarray}}
\newcommand{\eea}{\end{eqnarray}}
\newcommand{\ba}{\begin{array}}
\newcommand{\ea}{\end{array}}
\newcommand{\nn}{\nonumber}
\makeatletter \@addtoreset{equation}{section}

\makeatother
\begin{document}
\baselineskip 18pt%

\begin{titlepage}
\hfill%
\vbox{\halign{#\hfil \cr SUT-P-06-5B \cr IPM/P-2006/071 \cr
hep-th/0701046v3 \cr January 06,2007 \cr}}
\vspace*{12mm}%
\begin{center}
{\Large{\bf Matrix Theory for the DLCQ of\\ Type IIB String Theory on the AdS/Plane-wave}}%
\vspace*{10mm}

{\bf Mahdi Torabian}%

{\sl Institute for Studies in Theoretical Physics and Mathematics (IPM)\\
P.O.Box 19395-5531, Tehran, Iran \&\\
Department of Physics, Sharif University of Technology\\
P.O.Box 11365-9161, Tehran, Iran}\\
{\tt E-mail:mahdi@theory.ipm.ac.ir}%
\vspace*{15mm}

{\bf Abstract}%
\end{center}

We propose a recipe to construct the DLCQ Hamiltonian of type IIB
string theory on the AdS (and/or plane-wave) background. We consider
a system of $J$ number of coincident unstable non-BPS D0-branes of
IIB theory in the light-cone gauge and on the plane-wave background
with a compact null direction, the dynamics of which is described by
the world-line $U(J)$ gauge theory. This configuration suffers from
tachyonic instabilities. Having instabilities been cured through the
process of open string tachyon condensation, by expanding the theory
about true minima of the effective potential and furthermore taking
low energy limit to decouple the heavy modes, we end up with a
0+1-dimensional supersymmetric $U(J)$ gauge theory, a Matrix Theory.
We conjecture that the Hamiltonian of this Matrix Theory is just the
DLCQ Hamiltonian of type IIB string theory on the AdS or
equivalently plane-wave background in a sector with $J$ units of
light-cone momentum. We present some pieces of evidence in support
of the proposal.

\vspace*{5mm}\ 
\end{titlepage}
\tableofcontents
\section{Introduction and Motivation}
Over the course of time and through revolutionary progress, the
question of what is string theory, has been changed to what is
M-theory. M-theory has not been fully formulated. The only proposal
yet for its quantum definition, {\it i.e.} Matrix theory, is based
on describing the theory in terms of its Hamiltonian in the discrete
light-cone quantization (DLCQ), which is a simple $J$-body quantum
mechanical Hamiltonian ${\bf H}_J^{\rm DLCQ}$.

There exists a 1-parameter family of 0+1-dimensional $U(J)$ gauge
theories in the form of supersymmetric Matrix quantum mechanics
which is conjectured to give the DLCQ description of M-theory in the
sector with $J$ units of light-cone momentum on the plane-wave
background, called the BMN Matrix theory \cite{BMN}. The action is
basically describing or described by the dynamics of $J$ BPS
D0-branes of type IIA, which are gravitons from the eleven
dimensional viewpoint \cite{SS}. The parameter $\mu$, coming from
the background, characterizes a homotopy where at its starting point
$\mu=0$, sits the seminal BFSS Matrix theory as the DLCQ of M-theory
on the flat background which is the 0+1-dimensional SYM theory with
16 supercharges \cite{BFSS}. The extra adjustable parameter makes
the theory fascinatingly tractable which is useful in taking limits
or performing perturbative expansion. It removes all the flat
directions in the effective potential and puts barriers which makes
it possible to distinguish single and multi-particle states. It also
implies discrete spectrum and isolated set of normalizable vacua in
the form of fuzzy spheres which in the M-theory (continuum) limit
$J\rightarrow\infty$, have interpretation of spherical M2-brane
giant gravitons \cite{giant-graviton}.

Albeit the DLCQ procedure was used to give a non-perturbative and
second-quantized definition of M-theory, in principle it can be
employed to define any string theory by giving a recipe how to
construct the DLCQ Hamiltonian, ${\bf H}_J^{\rm DLCQ}$
\cite{DLCQ-Sen}. Furthermore, it is believe that the idea of DLCQ is
most natural in the context of Matrix theory \cite{Susskind}.

The DLCQ is in fact the light-cone quantization of a theory while
the null direction is compactified. In the light-cone frame the
basic coordinates are $X^+,X^-,X^I$ with conjugate momenta $P^-,
p^+, P^I$. The effect of compactification $X^-\sim X^-+ 2\pi R_-$ is
to discretize the spectrum of its conjugate momentum $p^+ = J/R_-$.
It provides a convenient IR regulator for the theory. Moreover, if
$p^+$ is positive and conserved, the Fock space of the system splits
into an infinite number of superselection sectors characterized by
$J$. In each sector states with $p^+ = J/R_-$ can have at most $J$
particles in them each carrying at least one unit of momentum. Thus,
the DLCQ of a theory in a given sector reduces to a quantum
mechanics with fixed number of particles \cite{DLCQ-Sen,DLCQ}. Hence
if we have a theory governing the dynamics of these $J$ partons, by
definition, it gives the dynamics of the original theory in the
sector $J$ of its DLCQ. It has been argued that the dynamics of
these $J$ partons is mapped into the dynamics of $J$ KK modes of the
same theory compactified on a space-like circle of radius $R$ once
the limit $R\rightarrow 0$ is taken \cite{DLCQ-Sen}. What would be
needed, inspired by some theory of partons, is to give a recipe for
constructing this $J$-body quantum mechanical Hamiltonian.

The aim of this note is to propose a recipe to construct the DLCQ
Hamiltonian ${\bf H}^{\rm DLCQ}_J$ of type IIB string theory on the
AdS/plane-wave backgrounds. In the light of the above facts and
inspired by the web of dualities, we consider a system of $J$ number
of non-BPS D0-Branes of type IIB string theory in the light-cone
gauge on the plane-wave background with a compact null direction,
the dynamics of which is described by the world-line $U(J)$ gauge
theory.

Besides usual stable BPS D$p$-branes for odd $p$, spectrum of type
IIB theory contains unstable non-BPS D$p$-branes with even $p$
\cite{non-BPS}. They do not carry any net RR charge, so because of
their tension they are unstable to decay to lighter neutral closed
string states. Fluctuations of the brane are captured by open
strings ending on it, so it is a natural expectation that
instability of the brane can also be encoded in open strings. In
fact, unstable branes contain tachyon in the open string spectrum on
their world-volume that is not removed by the usual GSO projection.
The tachyonic mode is a Higgs type excitation which spontaneously
acquires a vacuum expectation value and develops a stable state.

Although our system of interest is unstable, we show it could be
tractably stabilized. Indeed, the collection of non-BPS D0-branes in
the plane-wave background develops a non-trivial effective potential
for transverse coordinates and tachyon field which demonstrates
tachyonic instabilities. In fact this setup suffers from two sorts
of tachyonic instabilities which originate from each non-BPS
D0-brane separately and collectively a bunch of them in the presence
of the RR flux. They are respectively represented by tachyon field
in the spectrum of an open string connecting a brane to itself and
tachyonic modes of off-diagonal elements of the transverse scalars
which represent open strings stretching between the branes. These
instabilities are cured through the process of open string tachyon
condensation and the system dynamically stabilizes and falls into
true minima of the effective potential. By expanding the theory
about the true minima and taking a low energy limit to decouple the
heavy fluctuations, we end up with a 1-parameter family of
0+1-dimensional supersymmetric $U(J)$ gauge theories, a Matrix
theory.

We propose that the Hamiltonian of this Matrix Theory is just the
DLCQ Hamiltonian of type IIB string theory on the AdS or
equivalently plane-wave backgrounds in a sector with $J$ units of
light-cone momentum. This Matrix theory provides a non-perturbative
and second-quantized formulation of type IIB string theory and would
give an alternative definition and a better understanding of this
theory.

There is yet another proposal to construct the DLCQ Hamiltonian of
type IIB string theory on the AdS/plane-wave backgrounds. It has
been noted that the right probe in the presence of RR flux is
spherical D3-brane giant gravitons. Thus it is a natural expectation
that quantum completion of type IIB on this background comes from
quantum D3-brane theory. In \cite{TGMT} a prescription is given to
regularize light-cone D3-brane theory via discretizing it
world-volume. It has been conjectured that time-independent
volume-preserving diffeomorphism is isomorphic to $U(\infty)$ and
can be truncated to $U(J)$. Thus DLCQ of a D3-brane on the
plane-wave background leads to a 0+1-dimensional supersymmetric
$U(J)$ gauge theory, a Matrix theory, which is proposed to be the
DLCQ of type IIB string theory in the sector with $J$ units of
light-cone momentum on the AdS/plane-wave backgrounds. It is named
tiny graviton Matrix theory (TGMT). we will see that these two
proposal exactly coincide.

This paper is organized as follows; in section 2 we study the system
of interest, a system of $J$ number of coincident non-BPS D0-branes
of type IIB string theory on plane-wave background with a null
compact direction. Having fixed gauge redundancies, we derive
conjugate momenta of the system. In section 3 we analyze
stabilization processes via open string tachyon condensation. Then
by expanding the theory about true minima accompanied with taking
low energy limit we are led to a physically well-defined Matrix
theory. Having this theory at hand, in section 4 we solve its
equations of motions and derive various solutions. Taking all into
account, we write the light-cone Hamiltonian of this Matrix Theory
and proposed to be the DLCQ Hamiltonian of type IIB string theory on
the AdS/plane-wave backgrounds. In section 5 we elaborate on our
proposal. We introduce yet another proposal of Matrix theory for
DLCQ of type IIB string theory on AdS/plane-wave. It will be shown
that these two proposals are exactly the same and thus it is a
natural expectation that they share in evidence. We present some
pieces of evidence in support of the proposal. They are basically
based on symmetry structure of the theory, its vacuum structure,
spectrum of fluctuations about the vacua and studying the BPS
states. For physical applications, we will show how this Matrix
theory behaves under the string theory or decompactification limit.
Finally, in the last section we conclude and based on present ideas
give an outlook for future works.
\section{The Setup}
In this section we consider a system of $J$ number of coincident
non-BPS D0-brane of type IIB string theory on the maximally
supersymmetric ten dimensional plane-wave background\footnote{In
\cite{Ramgoolam}, system of non-BPS D0-branes on the
non-supersymmetric flat background with 4-form RR potential has been
studied.} which has a null circle.

\subsection{The action}%
Dynamics of massless bosonic and fermionic as well as tachyonic
modes of the non-BPS D$p$-branes of type II is suitably described by
generalized DBI and CS world-volume actions
\cite{Sen-action,RR-coupling,Bergshoeff,Garousi,Kluson}.
Supersymmetry is of course not manifest due to presence of the
tachyon and is regarded as spontaneously broken. Furthermore the
relationship between BPS and non-BPS branes implies that T-duality
must also hold for non-BPS branes, so we require the action to be
T-dual covariant. The low energy effective world-volume action has
been obtained as a function of the tachyonic and massless mode where
all the infinite massive mode is integrated out.

The non-Abelian extension of this action for a system of non-BPS
D$p$-branes has also been worked out in the literature. Demanding
the $U(J)$  invariance and T-duality covariance\footnote{It is also
interesting to require its consistency under S-duality for IIB
case.} it takes a unique form. Indeed the guiding principle in
constructing such an action is its consistency with the rules of
T-duality \'{a} la Myers \cite{Myers}\footnote{See also the earlier
works \cite{VanRaamsdonk}.} and the starting point is the action of
space filling non-BPS D9-brane of type IIA. Dynamics of bosonic
sector is given by
\cite{Garousi,Janssen-Meessen}%
\be\label{starting-action}\begin{split} {\cal S} = &-\beta T_9\int
{\rm d}^{10}\sigma{\rm Tr} \Big[e^{-\Phi}V(T)
\Big|-\det\big(G_{\mu\nu}+B_{\mu\nu}+2\pi\alpha'F_{\mu\nu}+2\pi\alpha'{\cal
D}_\mu T{\cal D}_\nu T\big)\Big|^{1/2}\Big] \cr &+ \beta Q_9\int\sum
C_n\wedge {\cal
D}{\rm Tr}\big[T\wedge e^{2\pi\alpha'F+B}\big], \end{split}\ee%
where $\beta$ is a constant, $G$ and $B$ are the metric and the
Kalb-Ramond field of the bulk respectively and $F$ is the field
strength of the gauge field on the brane. $T$ is the tachyon field
with the tachyon mass $m^2=-1/2\alpha'$ and its potential is
$V(T)=e^{2\pi\alpha'm^2T^2}$ which is zero at the minimum
$V(T_0)=0$. Tension and charge of the brane is $ T_9 = Q_9 =
l^{-9}_s$. Covariant derivative is defined as ${\cal D}_{\mu} =
\partial_\mu + i[{\cal A}_\mu,\cdot\ ]$,\ $\mu=0,1,\dots 9$.
The gauge field ${\cal A}_\mu$ and $T$ sit in the adjoint
representation of the gauge group. Trace is taken completely
symmetrized between all non-Abelian expressions in $U(J)$
representation. Determinant is taken over $SO(10)$ representations.

In order to construct the non-Abelian low energy effective action of
$J$ non-BPS D0-branes, we apply T-duality transformation rules to
D9-brane in 9 longitudinal directions. The
final form of the action is%
\be\begin{split} {\cal S} = &-\beta T_0 \int {\rm d}\tau\ {\rm
Tr}\Big[e^{\phi}V(T)\det\,^{1/2} Q^I\,_J\Big|-\Big({\cal
P}\big[G_{00} + G_{0I}(Q^{-1}-\delta)^I\, _K G^{KJ}G_{J0}\big] +
T_{00}\Big)\Big|^{1/2}\Big] \cr &+\beta Q_0\, \int {\rm d}\tau\ {\rm
Tr}\Big[{\cal P}\big[e^{i/(2\pi\alpha') \iota_X\iota_X}\ \sum_n
C_n\wedge e^B.\ \big(-i/(2\pi\alpha')^{1/2}
\overleftarrow{\iota}_{[X,T]} + \wedge {\cal D}T\big)\big]\Big].
\end{split}\ee%
Fluctuations of the transverse collective coordinates is manifested
as another potential and is encoded in the matrix $Q$%
\be Q^I\,_J = \delta^I\,_J - \frac{i}{2\pi\alpha'}[X^I,X^K]E_{KJ}
- \frac{1}{2\pi\alpha'}[X^I,T][X^K,T]E_{KJ}, \ee%
and contribution of tachyonic modes to the dynamics is entering
through matrix $T_{00}$%
\be\begin{split}T_{00} &= \big(2\pi\alpha' -
[X^I,T](Q^{-1})_{IJ}[X^J,T]\big){\cal D}_0T{\cal D}_0T \cr &-
iE_{0I}(Q^{-1})^I\,_J[X^J,T]{\cal D}_0T - i{\cal
D}_0T[X^I,T](Q^{-1})_I\,^JE_{J0} \cr &- i{\cal
D}_0X^I(Q^{-1})_{IJ}[X^J,T]{\cal D}_0T - i{\cal
D}_0T[X^I,T](Q^{-1})_{IJ}{\cal D}_0X^J, \end{split}\ee%
where we have defined $E_{\mu\nu}\equiv G_{\mu\nu}+B_{\mu\nu}$ with
which we raise or lower indices. $E^{IJ}$ denotes the inverse of
$E_{IJ}$. $I$'s are transverse indices to the D0-branes. Note that
there is no contribution from field strength, and the gauge field
contributes only through covariant derivative. Also, we use
covariant derivatives in pulling back the bulk fields. The potential
$\det^{1/2} Q$ is equal to one for the Abelian case.

\subsubsection*{The plane-wave background}%
We are interested in the ten dimensional plane-wave background which
is maximally supersymmetric and $\alpha'$-exact solution of
the IIb supergravity, specified by%
\bse\label{background}\begin{align}
ds^2 = -2 dX^+ dX^-  -&\mu^2X^I X^I{(dX^+)}^2 + dX^I dX^I\\
F_{+ijkl} = 4\frac{\mu}{g_s}\ \epsilon_{ijkl} &\quad,\quad
F_{+abcd}= 4\frac{\mu}{g_s} \ \epsilon_{abcd} \\
e^\phi = g_s &= \rm{constant}. \end{align}\ese%
The RR 4-form potential in a gauge which maintains translational
symmetry along $X^+$ reads as%
\be C_{+ijk} = - \frac{\mu}{g_s}\ \epsilon_{ijkl}X^l \quad,\quad
C_{+abc} = -\frac{\mu}{g_s}\ \epsilon_{abcd}X^d. \ee%
This background has a globally defined light-like Killing vector
$\partial/\partial X^-$ and one dimensional light-like causal
boundary. The parameter $\mu$ whose value is arbitrary has dimension
of energy. For a detailed discussion regarding this background see
\cite{plane-wave-review}.

\subsubsection*{The gauge fixing}%
Next we fix gauge redundancies. There are two of them; The first one
is 1-dimensional diffeomorphism along the world-line. Due to
symmetries of the plane-wave background and in particular
translational symmetries along the light-like directions, fixing the
light-cone gauge will considerably simplify the action. We do this
by identifying world-line time with one of the light-come
coordinates,
using and fixing worldline reparametrization%
\be \tau \sim X^+. \ee%
$X^+$ and $X^-$ is no longer dynamical variables. $X^-$ is a cyclic
coordinate, so its conjugate momentum $p^+$ is a constant of motion.

The second gauge redundancy is the internal gauge symmetry. We use
temporal gauge
\be{\cal A}_0 = 0\ee%
to fix it. Then, we impose its equation of motion as constraint on
the system or physical condition for the states.

\subsection{The conjugate momenta of the system}%
Putting back the background fields in the action and fixing the
light-cone gauge we uncover the effective action of this
configuration as%
\be\label{action} {\cal S}[X^+,X^-,X^I,{\cal A}_0,T] = \int {\rm
d}X^+\ \big(L_{DBI}+L_{CS}\big), \ee%
\be\begin{split} L_{DBI} = {\rm Tr}\Big[& -\frac{1}{l_sg_s}\ V(T)
\det\,^{1/2}Q \times  \cr & \times \Big| \mu^2X_I^2(\partial_0X^+)^2
+ 2\partial_0X^+\partial_0X^- - {\cal D}_0X_I (Q^{-1})^{IJ} {\cal
D}_0X_J \cr &- {\cal D}_0 T\big(2\pi\alpha' -
[X_I,T](Q^{-1})^{IJ}[X_J,T] \big){\cal D}_0 T \cr & + 2i\big({\cal
D}_0 X_I (Q^{-1})^{IJ}[X_J,T]{\cal D}_0 T + {\cal D}_0
T[X_I,T](Q^{-1})^{IJ} {\cal D}_0 X_J \big)\Big|^{1/2}\Big],
\end{split} \nn\ee%
\be\hspace{-9mm} L_{CS} = \frac{\mu}{(2\pi)^3g_sl^4_s}{\rm
Tr}\Big[[X^j,X^i]\epsilon^{ijkl}X^k[X^l,T] +
[X^b,X^a]\epsilon^{abcd}X^c[X^d,T]\Big]\partial_0X^+, \nn\ee%
where the matrix $Q$ is of the form%
\be Q_{IJ} = \delta_{IJ} - \frac{i}{2\pi\alpha'}[X_I,X_J] -
\frac{1}{2\pi\alpha'}[X_I,T][X_J,T], \ee%
and $Q^{-1}$ is its inverse. $Q$ is an $8\times 8$, as well as a
$J\times J$ matrix. The $\det Q$ is however, only on $8\times 8$
indices and can be expanded as%
\be\label{Q-deteminant} \det Q =
\sum_{n=0}^\infty\frac{1}{n!}\bigg({\rm
Tr}\sum_{m=1}^\infty\frac{(-1)^{m+1}}{m}\sum_{k=0}^mC_k^m(-1)^m
\chi\Big[[X,X]^k\big([X,T][X,T]\big)^{m-k}\Big]\bigg)^n, \ee%
where we have to consider all possible permutations $\chi$ of
commutators, because they are matrices rather than $c$-numbers. Now
having the action at hand, we derive conjugate momenta associated
with non-dynamical $X^+, X^-$ and dynamical variables $X^I, T, {\cal
A}_0$.

The Light-cone momentum, momentum conjugate to $X^-$ is defined as
$p^+ = -P_- = -\partial{\cal L}/\partial\dot X^-$, and can
explicitly written as%
\be\begin{split} p^+ = \frac{1}{Jl_sg_s} {\rm Tr}\Big[ &V(T)
\det\,^{1/2}Q \times \cr &\times\Big( \mu^2X_I^2 + 2
\partial_0X^- - {\cal D}_0X_I (Q^{-1})^{IJ} {\cal D}_0X^J \cr & -
{\cal D}_0 T\big(2\pi\alpha' - [X_I,T](Q^{-1})^{IJ}[X^J,T]
\big){\cal D}_0 T \cr & + 2i\big({\cal D}_0 X_I
(Q^{-1})^{IJ}[X_J,T]{\cal D}_0 T + {\cal D}_0 T[X_I,T](Q^{-1})^{IJ}
{\cal D}_0 X_J \big)\Big)^{-1/2}\Big],
\end{split}\ee%
As proposed in the introduction, the light-cone momentum should be
distributed among $J$ D0-branes, strictly speaking each D0-brane
carries one unit of light-cone momentum. Hence, the eigenvalue of
this operator should be somehow discretized. It can directly be
obtained by compactifying the longitudinal light-cone coordinate
$X^- \sim X^-+2\pi R_-$ which results in discretized eigenvalues for
corresponding conjugate momenta as%
\be p^+=J/R_-\ .\ee%
We have chosen the radius of compactification $R_-$ in such a way
that $J$ appears in the numerator.

Transverse momenta, momenta conjugate to $X_I$ are $P_I =
\partial{\cal L}/\partial\dot X^I$
\be P^I = \frac{1}{2}p^+ \Big[{\cal D}_0X_J (Q^{-1})^{IJ} +
(Q^{-1})^{JI}{\cal D}_0X_J - 2i (Q^{-1})^{IJ}[X_J,T]{\cal D}_0T -
2i{\cal D}_0T [X_J,T](Q^{-1})^{JI}\Big]. \ee%

Momentum conjugate to the tachyon field is $P_T = \partial{\cal
L}/\partial\dot T$,%
\be\begin{split} P_T = 2\pi\alpha' p^+ \Big[ {\cal D}_0 T & +
2\pi\alpha'{\cal D}_0 T[X_I,T](Q^{-1})^{IJ}[X_J,T] \cr &-
2\pi\alpha'\big([X_I,T](Q^{-1})^{IJ}{\cal D}_0X_J - {\cal
D}_0X_I(Q^{-1})_{IJ}[X_J,T] \big) \Big]. \end{split}\ee%

The light-cone Hamiltonian, that is the momentum conjugate to $X^+$,
is ${\bf H} = P^- = -p_+ = -\partial{\cal L}/\partial\dot X^+$.
Explicitly,%
\be\begin{split} J{\bf H} = p^+{\rm Tr}\bigg[&
\frac{1}{2(p^+l_sg_s)^2}\ V^2(T) \det Q + \frac{1}{2} \mu^2 X_I^2 +
\frac{1}{2} \partial_0X_I (Q^{-1})^{IJ} \partial _0X_J \cr & +
\frac{1}{2} \partial_0 T\Big( 2\pi\alpha' -
[X_I,T](Q^{-1})^{IJ}[X_J,T]\Big) \partial_0 T \cr & - i
\Big(\partial_0 X_I (Q^{-1})^{IJ}[X_J,T]\partial_0 T + \partial_0
T[X_I,T] (Q^{-1})^{IJ} \partial_0 X_J\Big)\bigg] \cr
&\hspace{-11mm}-{\rm Tr}\bigg[
\frac{\mu}{(2\pi)^3g_sl^4_s}\Big([X^j,X^i]\epsilon^{ijkl}X^k[X^l,T]
+ [X^b,X^a]\epsilon^{abcd}X^c[X^d,T]\Big)\bigg]. \end{split}\ee%

Finally, momentum conjugate to the gauge field is zero $P_{{\cal
A}_0} = 0$. We fix this gauge redundancy in
temporal gauge ${\cal A}_0 = 0$ and impose its equation of motion%
\be \Phi = \frac{\partial{\cal L}}{\partial{\cal A}_0} =
\Big[X_I,\frac{\partial{\cal L}}{\partial {\cal D}_0X_I}\Big] +
\Big[T,\frac{\partial{\cal L}}{\partial {\cal D}_0T}\Big] =
[X^I,P_I] + [T,P_T],\ee%
as the Gauss law constraint on the system.

The light-cone Lagrangian is defined through Legender
transformation%
 \be {\bf L} = P_-\dot X^- +
P_I\dot X^I + P_T\dot T- {\bf H} - {\cal A}_0 \Phi, \ee%
and explicitly is%
\be\begin{split} J{\bf L} = p^+{\rm Tr}\bigg[& \frac{1}{2} {\cal
D}_0X_I (Q^{-1})^{IJ} {\cal D}_0X_J - \frac{1}{2(p^+l_sg_s)^2}V^2(T)
\det Q - \frac{1}{2} \mu^2 X_I^2 \cr & + \frac{1}{2} {\cal D}_0
T\Big( 2\pi\alpha' + [X_I,T](Q^{-1})^{IJ}[X_J,T]\Big) {\cal D}_0 T
\cr & + i \Big({\cal D}_0 X_I (Q^{-1})^{IJ}[X_J,T]{\cal D}_0 T +
{\cal D}_0 T[X_I,T] (Q^{-1})^{IJ} {\cal D}_0 X_J\Big)\bigg] \cr
&\hspace{-11mm}+{\rm Tr}\bigg[ \frac{\mu}{(2\pi)^3g_sl^4_s}
\Big([X^j,X^i]\epsilon^{ijkl}X^k[X^l,T] +
[X^b,X^a]\epsilon^{abcd}X^c[X^d,T]\Big)\bigg]. \end{split}\ee%
Having introduced our system of interest, in the following section
we continue to study its dynamics.
\section{Stabilization of the System}
We now analyze stabilization of the theory through the processes of
open string tachyon condensation. In order to analyze the dynamics,
we rewrite the action \eqref{action} as ${\cal S}= {\cal
S}_{kin.}+{\cal S}_{pot.}$ where%
\be {\cal S}_{pot.} = -\int{\rm d}X^+ {\bf V}(X,T,{\cal A}_0). \ee%
The collection of $J$ non-BPS D0-branes in the plane-wave background
develops a non-trivial effective potential ${\bf V}$. It is a
functional of $U(J)$ matrix valued fields $X^I, T, {\cal A}_0$ and
possible matrix commutators of them $[X,X], [X,T], [{\cal A}_0,X],
[{\cal A}_0,T]$ as well as a set of parameters $\alpha', g_s, \mu,
R_-, g_s$. It
reads%
\be\label{effective-potential}\begin{split} {\bf V}(X,T,{\cal A}_0)
= R_-{\rm Tr}\bigg[ &\frac{1}{2} \left(\frac{\mu}{R_-}\right)^2
X_I^2 \cr + &\frac{1}{2(Jl_sg_s)^2}V^2(T)\det Q \cr -
&\frac{1}{2R_-^2} {\cal D}_0 T [X_I,T](Q^{-1})^{IJ}[X_J,T]{\cal D}_0
T \cr - &\frac{i}{R_-^2} \big({\cal D}_0 X_I
(Q^{-1})^{IJ}[X_J,T]{\cal D}_0 T + {\cal D}_0 T[X_I,T] (Q^{-1})^{IJ}
{\cal D}_0 X_J\big) \cr - &\frac{\mu}{(2\pi)^3R_-g_sl^4_sJ}
\Big([X^j,X^i]\epsilon^{ijkl}X^k[X^l,T] +
[X^b,X^a]\epsilon^{abcd}X^c[X^d,T]\Big)\bigg]. \end{split}\ee%

The first term, the mass term of transverse coordinates, comes from
the metric through DBI part of the action and is due to the
back-reaction of the 5-form RR flux in the background. The rest
(except for trivial part of the $\det Q$ which gives the mass term
of the tachyon) are due to collective behavior of D0-branes as is
manifested in the commutators of the fields. The next three terms
are the potential for fluctuations of the transverse coordinates and
the tachyon coming also via DBI action. The last term originates
from the coupling of the transverse/tachyon fluctuations to the
4-form gauge potential through CS part of the
action\footnote{Throughout this note we are using temporal gauge for
the gauge field, as ${\cal A}_0 = 0$}.

This system suffers from two sorts of tachyonic instabilities. One
is due to tachyon field in the spectrum of open string connecting a
non-BPS D0-brane to itself. It is encoded in (diagonal part of) the
$J\times J$ matrix $T$. It gives further contribution to the
effective potential through $[X,T]$ terms besides the $V(T)$ term.
The other instability stems from tachyonic modes of the off-diagonal
modes of the scalars $X$'s. They are agents of coupling of
transverse fluctuations (represented by open strings stretching
between D0-branes) to the bulk RR flux\footnote{See similar analysis
for the BPS D0-branes of type IIA theory in the presence of 4-form
flux has been elaborated in detail in \cite{Kimura}.}. It gives
contribution via $[X,X]$ and $[X,T]$ commutators.

\subsection{Open string tachyon condensation}
Eventually, upon the process of open string tachyon condensation,
this system non-trivially stabilizes and falls into true extrema of
the effective potential. The tachyon field as well as matrix
commutators as agents of instabilities spontaneously acquire non-zero expectation values%
\be \langle T\rangle\neq 0,\qquad \langle [X,X]\rangle\neq 0,
\quad \langle [X,T] \rangle\neq 0 \ee%
and all the modes condensate at the minima of the potential. In the
next section we will see that in fact there is a set of minima and
study the theory at these points. It is considerable noting that
commutators are not quantum commutation relations but just matrix
commutators and so is the expectation values, they are statistical
(average values) not quantum mechanical.

In the absence of RR flux the preferred configuration is the one
with $[X,X]=0$ and $[X,T]=0$. It defines a moduli space on which $X$
and $T$ matrices are simultaneously diagonalizable. Before the
process of open string tachyon condensation, it has interpretation
of positions for the non-BPS D0-branes. However, in the presence of
the flux, they no longer commute and the classical interpretation of
D0-brane positions breaks down. There is a fuzziness in description
of their positions and we define the mean-square value of the $i$th
coordinate to be $\langle X_i^2\rangle\ \sim 1/J$ when averaged over
$J$ D0-branes. Upon tachyon condensation, the values of commutators
become specific in such a way that extremize the effective
potential.

\subsection{Expansion about true minima}
At the minima of the potential there is a stable theory, the
potential of which can by obtained be expanding the above effective
potential about each of the minima in the form of a harmonic
oscillator potential. As we discussed there are two different
physical processes which are governed by $[X,X]$, $[X,T]$ as well as
$T$ in the interaction part of the above action. They can be seen as
orthogonal directions in the graph of effective potential (there is
yet another orthogonal direction related to mass term of $X$,
although disjoined from the others). We extremize the potential with
respect to all of them and expand appropriately about the minima of
them. To have a potential of the form of harmonic
oscillator, we keep terms second-order in derivatives, as%
\be {\bf V} = V_0 + \frac{\delta^2{\bf
V}}{\delta\Phi\delta\Phi}\Big|_{\Phi=\Phi_0} (\Phi-\Phi_0)^2 + \cdots\ , \ee%
where $\Phi$ could be either of $X^I, [X^I,X^J][X^K,T]$ or $T$. As
we will se in the next section, they are really true minima and not
saddle points.

\subsection{The decoupling limit}
Furthermore, we decouple the heavy modes and fluctuations about
these minima, by taking low energy limit
$\alpha'\sim\epsilon\rightarrow 0$ while keeping the physical
light-cone momentum $\mu p^+\alpha'$, string coupling $g_s$, $J$ and
$R_-$ fixed. To be consistent, we send dimensionful parameter of the
background $\mu\sim\epsilon^{-1}\rightarrow\infty$. Prior to taking
the limit, we rescale parameters properly to bring
them in energy dimension as%
\be X\rightarrow\alpha' X,\quad
\alpha'^{-1}[X,X]\rightarrow\alpha'[X,X],\quad
\alpha'^{-1/2}[X,T]\rightarrow\alpha'^{1/2}[X,T], \ee%
and leave $T$ unchanged. In the low energy regime, we just keep
terms in the determinant of $Q$ to order ${\cal O}(\alpha'^2)$ and
higher order terms decouple from the theory.

It is a non-trivial possibility to take a limit where $\langle
T\rangle$ remains finite while the fluctuations of $T$ about it
become very massive and decouple. From now on we refer to matrix of
vacuum expectation values of tachyon as $\langle T\rangle = {\cal
T}$ which can take various values, and of course matrix form.
Furthermore, we set $P_T =0 $ and $V(T)$ becomes a constant matrix.

\subsection{The action for the stabilized phase of the system}
Finally having done all this, the harmonic oscillator potential
about the true minima takes this form%
\be\begin{split} {\bf V} = R_-{\rm Tr}\bigg[ &\frac{1}{2}
\left(\frac{\mu}{R_-}\right)^2 X_I^2 \cr +
&\frac{1}{2(Jl_sg_s)^2}V^2({\cal
T})\chi\Big[[X^I,X^J][X^K,X^L][X^M,{\cal T}][X^N,{\cal T}]\Big] \cr
- &\frac{\mu}{(2\pi)^3R_-g_sl^4_sJ}
\Big([X^j,X^i]\epsilon^{ijkl}X^k[X^l,{\cal T}] +
[X^b,X^a]\epsilon^{abcd}X^c[X^d,{\cal
T}]\Big)\bigg]. \end{split}\ee%
For later use, we also derived the light-cone Lagrangian of this
Matrix theory. It reads%
\be\begin{split} {\bf L} = R_-\ {\rm Tr}&\Big[ \frac{1}{2R_-^2}\
({\cal D}_0X_I)^2 - \frac{1}{2}\left(\frac{\mu}{R_-}\right)^2X_I^2
\cr &- \frac{1}{2.3!\ l_s^8g_s^2J^2}\ V^2({\cal
T})[X^I,X^J,X^L,{\cal T}][X^I,X^J,X^L,{\cal T}] \cr &
+\frac{\mu}{3!l_s^4R_- g_sJ}\left( \epsilon^{i j k l} X^i [X^j, X^k,
X^l,{\cal T}]+ \epsilon^{a b
c d} X^a [ X^b, X^c, X^d ,{\cal T}] \right)\Big], \end{split}\ee%
where the 4-commutator is defined as%
\be\begin{split} [A,B,C,{\cal T}] = \frac{1}{4!}\Big(&[A,B][C,{\cal
T}]- [A,C][B,{\cal T}] + [A,{\cal T}][B,C] \cr +&[C,{\cal T}][A,B] -
[B,{\cal T}][A,C] + [B,C][A,{\cal T}]\Big),\end{split}\ee%
and we have used the identity%
\be \epsilon^{ijkl}X^iX^jX^kX^l =
\frac{1}{4!}\epsilon^{ijkl}[X^i,X^j,X^k,X^l] =
\frac{1}{2.4!}\epsilon^{ijkl}[X^i,X^j][X^k,X^l]. \ee%

The action%
\be {\cal S} = \int dX^+ {\bf L} \ee%
gives the dynamics of a physically well-defined and stable
0+1-dimensional $U(J)$ gauge theory, a Matrix theory. Furthermore,
as is noted before, the supersymmetry in the unstable theory with
tachyon field is regarded as spontaneously broken. Upon tachyon
condensation the supersymmetry is restored and we end up with a
supersymmetric gauge theory. If, besides tachyonic and massless
bosonic sector, we had also added massless fermionic sector to the
action we started with \eqref{starting-action}, the above action for
the stable theory would be manifestly supersymmetric.
\section{The Proposal}

\subsection{Equations of motion}%
We now study this Matrix theory by solving its equation of motion
for dynamical fields. We will see that the solutions are in the form
of concentric fuzzy 3-spheres, which form various vacuum
configurations of the theory.

The equation of motion of transverse adjoint scalars $X^i$
(and similarly for $X^a$) is%
\be D_0P_i - i\frac{\partial{\bf L}}{\partial D_0X^j}\frac{\partial
[{\cal A}_0,X^j]}{\partial X^i} - \frac{\partial{\bf
L}}{\partial[X^j,X^k,X^l,{\cal T}]}\frac{\partial [X^j,X^k,X^l,{\cal
T}]}{\partial X^i} - \frac{\partial{\bf L}}{\partial X^i} = 0,\ee%
which can be written as%
\be\begin{split} \partial_0P^i&+\frac{1}{2.3!g_s^2l_s^2J^2}\
\big[X^j,X^k,[X^i,X^j,X^k,{\cal T}],{\cal T}\big] \cr &+
\left(\frac{\mu}{R_-}\right)^2X^i - \frac{\mu}{3!l_s^4R_- g_sJ}\
\epsilon^{ijkl}[X^j,X^k,X^l,{\cal T}] =0. \end{split}\ee%

Although tachyon is condensed at the minima of the tachyon
potential, we can treat it as a variable and compute its equation of
motion, it reads as%
\be - \frac{\partial{\bf L}}{\partial[X^I,X^J,X^K,{\cal
T}]}\frac{\partial [X^I,X^J,X^K,{\cal T}]}{\partial
{\cal T}} - \frac{\partial{\bf L}}{\partial {\cal T}} = 0,\ee%
which can explicitly be written as%
\be\begin{split} &
\frac{1}{2.3!g_s^2l_s^8J^2}\Big(\big[X^I,X^J,X^K,[X^I,X^J,X^K,{\cal
T}]\big] + 4{\cal T}V^2({\cal T})[X^I,X^J,X^K,{\cal T}]^2\Big)\cr -&
\frac{\mu}{2.3!l_s^4R_- g_sJ}
(\epsilon^{ijkl}[X^i,X^j,X^k,X^l]+\epsilon^{abcd}[X^a,X^b,X^c,X^d])
= 0. \end{split}\ee%
Equation of motion of the tachyon can be interpreted as a constraint
on the system. Namely, having solved $X$ equations of motion, the
configurations should also satisfy $T$ constraint equation.

Equation of motion for ${\cal A}_0$ reads%
\be i[X^I,P_I] = 0, \ee%
which is the Gauss' law and should be satisfied by all the physical
configurations.

We are looking for static solutions $P^I=0$ of the equations of
motion for which the potential is extremum. Furthermore, we consider
a class of
solutions for which%
\be [X^i,X^j,X^a,{\cal T}] = [X^a,X^b,X^i,{\cal T}] = 0. \ee%
With this, equations mixed between $i$ and $a$ directions decouple.
Hence, the equations of motion reduce to (the same set of equation
holds for $X^a$)%

$X^i:$ \be\label{1st-equation}\begin{split}
&\frac{1}{2.3!g_s^2l_s^8J^2}\big[X^j,X^k,[X^i,X^j,X^k,{\cal
T}],{\cal T}\big] + \left(\frac{\mu}{R_-}\right)^2X^i \cr -&
\frac{\mu}{2.3!l_s^4R_-
g_sJ}\ \epsilon^{ijkl}[X^j,X^k,X^l,{\cal T}] =0 \end{split}\ee%

${\cal T}:$ \be\label{2nd-equation}\begin{split}
&\frac{1}{2.3!g_s^2l_s^8J^2}\Big(\big[X^i,X^j,X^k,[X^i,X^j,X^k,{\cal
T}]\big] + 8{\cal T}[X^i,X^j,X^k,{\cal T}]^2 \Big) \cr -
&\frac{\mu}{2.3!l_s^4R_- g_sJ} \big(\epsilon^{ijkl}[X^i,X^j,X^k,X^l]
\big) = 0 .\end{split}\ee%

\subsection{The anzats}
The solution to the $X$ equation of motion \eqref{1st-equation}, is%
\be\label{1st-def} [X^i,X^j,X^k,{\cal T}] = -\frac{\mu g_s}{R_-}\
\alpha'^2J\epsilon^{ijkl}X^l . \ee%
put it back in ${\cal T}$ equation of motion (constraint equation)
\eqref{2nd-equation}, consistency requires%
\be\label{1st'-def} [X^i,X^j,X^k,X^l] = \Big(\frac{\mu
g_s}{R_-}\ \alpha'^2J\Big)^2 \epsilon^{ijkl}{\cal T} , \ee%
together with%
\be\label{2nd-def} \sum_{i=1}^4 \delta^{ij}X^iX^j = \frac{\mu
g_s}{R_-}\ \alpha'^2J\ . \ee%
Consistency among equations \eqref{1st-def}-\eqref{2nd-def} requires
${\cal T}$ anticommutes with $X$ and furthermore, squares to ${\bf
1}$. Hermiticity and tracelessness together with the fact that
${\cal T}$ could become diagonalized, fix it to be a $J\times J$
matrix with equal number of +1 and -1 eigenvalues. More details
about realization of the solutions to matrix equations
\eqref{1st-def}-\eqref{2nd-def} is given in appendix
\ref{appendix-fuzzy-sphere}.

Solution to the equations of motions, equations \eqref{1st-def} and
\eqref{2nd-def}, define fuzzy 3-sphere (see appendix
\ref{appendix-fuzzy-sphere}),
the radius $R$ and fuzziness $l$ of which is defined as%
\be R_{fuzzy} = \left(\frac{\mu g_s}{R_-}J\right)^{1/2}\alpha'
\qquad;\qquad l = \left(\frac{\mu g_s}{R_-}\right)^{1/2}\alpha' . \ee%

Equations \eqref{1st-equation} and \eqref{2nd-equation} are matrix
equations and we have to take into account matrix form of $X$'s and
${\cal T}$. Hence, we must consider all the possible forms of the
matrices as well as coefficients which solve the above equations. If
one considers block-diagonal set of matrices $X$'s and ${\cal T}$
then the classical equations of motion for the blocks are separable.
One can think of these blocks as describing different matrix theory
objects which obeying classically independent equations of motions.
This gives an implication how Matrix theory can encode a
configuration of multiple objects \cite{Taylor}.

As we argued in previous section, there is a set of minima for the
effective potential at which the tachyons can be condensed and
spontaneously take nonzero vacuum expectation values and Matrix
form. These extrema corresponds to different configurations of
concentric fuzzy 3-spheres. Generically, a solution could be of the
following form when $J\times J$ matrices $X$ and ${\cal T}$ is
partitioned into $k$ blocks of size $J_i$ in
such a way that $\sum_{i=1}^{k} J_i = J$, as%
\begin{figure}[h]\be
\includegraphics[scale=1.26]{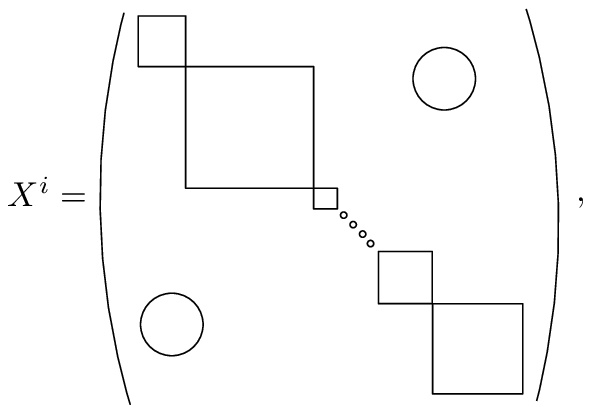}
\hspace*{2mm}
\includegraphics[scale=1.26]{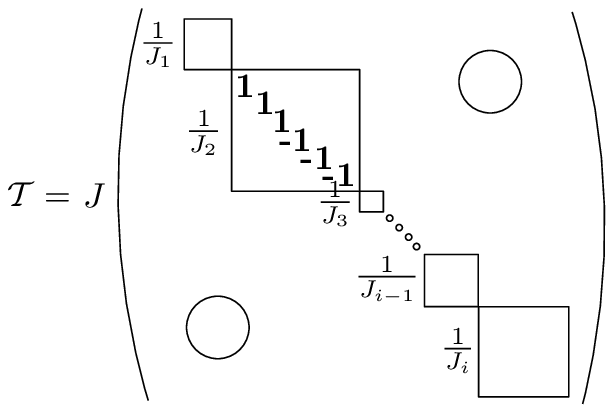}.
\ee\end{figure}%
\\ Each block in the matrix ${\cal T}$ has $J_i/2$ number +1 and
$J_i/2$ number -1 eigenvalues. $X$'s have the same pattern of
block-diagonalization as ${\cal T}$. They obey independent
equation of motion and consistency demands%
\be \sum_{i=1}^4
\delta^{ij}X^iX^j = \frac{\mu
g_s}{R_-}J_i\ . \ee%
for each block inside $X$'s. These solutions are in the form of $k$
concentric fuzzy 3-spheres. Actually, solutions can be classified
and labeled in terms of irreducible or reducible $J\times J$
representation of $spin(4)$ which respectively correspond to single
and multi concentric fuzzy 3-spheres of various radii extended in
$X^i$ and/or $X^a$ directions \cite{half-BPS}. Multi solutions are
related to all the possible partitions of $J$ and the radius of each
sphere is given by the
fraction of the total light-cone momentum $J_i$ it carries%
\be R_{fuzzy} = \left(\frac{\mu g_s}{R_-}J_i\right)^{1/2}\alpha' . \ee%
The fuzziness, which is characterized by the parameters of the
background and string coupling constant, is a unique parameter for
all the solutions and given by

\subsection{Statement of the proposal:\\Matrix theory Hamiltonian
as the DLCQ Hamiltonian} Taking all these into account, now we
rewrite bosonic part of the light-cone Hamiltonian of this Matrix
theory (once we rescale ${\cal
T}\to {\cal T}/J$) as%
\be\label{Hamiltonian-1}\begin{split} {\bf H} = R_-\ {\rm Tr}&\Big[
\frac{1}{2}\ P_I^2 + \frac{1}{2}\left(\frac{\mu}{R_-}\right)^2X_I^2
+ \frac{1}{2\cdot 3!l_s^8g_s^2}\ [X^I,X^J,X^L,{\cal T}]^2 \cr & -
\frac{\mu}{3!l_s^4R_- g_s}\left(
\epsilon^{ijkl}X^i[X^j,X^k,X^l,{\cal T}]+
\epsilon^{abcd}X^a[X^b,X^c,X^d,{\cal T}] \right)\Big],\
\end{split}\ee%
where now ${\cal T}$ is fixed to take only its vacuum values. This
Hamiltonian well defines a 0+1-dimensional $U(J)$ gauge theory in
the form of a Matrix theory. We propose that it is just the DLCQ
Hamiltonian, ${\bf H}_J^{\rm DLCQ}$ of type IIB string theory with
$J$ unit of light-cone momentum on the AdS/plane-wave backgrounds.

As is evident, it is just the bosonic sector of the DLCQ
Hamiltonian. The fermionic sector could similarly be derived. We
insist that upon tachyon condensation and decoupling of the extra
tachyonic degree of freedom, the supersymmetry is restored and the
stable theory is fully supersymmetric.

Another crucial point is that we have really expanded the theory
about the true minima of the effective potential
\eqref{effective-potential}, not its saddle points. It can be
justified by noting that all the terms in the potential of our
Matrix theory is positive definite%
\be\label{positive-definite}\begin{split} {\bf V} = R_-\ {\rm
Tr}\Big[ &\frac{1}{2}\bigl(\frac{\mu}{R_-} X^l + \frac{1}{3!l_s^4
g_s}\ \epsilon^{ijkl} [ X^i , X^j , X^k, {\cal T}]\bigr)^2 +
\frac{1}{4.3!l_s^8g_s^2} [X^i, X^j, X^a, {\cal T}]^2 \cr +
&\frac{1}{2}\bigl(\frac{\mu}{R_-} X^d+ \frac{1}{3!l_s^4 g_s}\
\epsilon^{abcd}[ X^a , X^b , X^c, {\cal T}]\bigr)^2 +
\frac{1}{4.3!l_s^8g_s^2} [X^i, X^a, X^b, {\cal T}]^2 \Big].\end{split}\ee%

\section{Considerations on the Proposal}
This section is devoted to tighten our conjecture by giving some
pieces of evidence in support of.
\subsection{Another proposal: Tiny Graviton Matrix Theory}
For completeness we review another proposal to construct the DCLQ
Hamiltonian of type IIB string theory on the AdS/planewave
backgrounds.

It has been noted that the right probe in the presence of RR flux is
spherical D3-brane giant gravitons \cite{giant-graviton}. It is a
natural expectation that quantum completion of type IIB on this
background comes from quantum D3-brane theory. Due to large amount
of unphysical degrees of freedom and gauge redundancies, it is
difficult to study its quantum mechanical properties. However, there
is an alternative formulation of the D3-brane as a gauge theory of
the volume-preserving transformations of the brane hypersurface
\cite{BSTT}. It is inspired by the fact that these transformations
are residual invariance of a D3-brane theory when formulated in the
light-cone gauge. Fixing the light-cone gauge fixes a part of the
diffeomorphism invariance of the action which mixes the world-volume
time with the world-volume spatial directions. The part of
diffeomorphisms which only act on the spatial directions are still
present and not fixed. This supersymmetric gauge theory provides a
convenient framework for study the quantum mechanical properties of
the D3-brane. It is possible to consider truncations of this gauge
theory by truncating the infinite harmonic expansion of the D3-brane
coordinates. At least for the D3-brane with the topology of a sphere
this can be done in such a way that the supersymmetry remains
preserved. These truncations lead to a class of matrix models in
supersymmetric quantum mechanics and D3-brane can be viewed as a
limiting case of them.

Consider the low energy effective action of the D3-brane%
\be {\cal S}_{D3} = - T\int d^{4}\sigma
e^{-\phi}\Big|-\det\big(P[G+B]+F\big)\Big|^{1/2} + Q\int C_{(n)},\ee%
where $T=Q=(2\pi)^3l_s^{-4}$. The world-volume theory of D3-brane
consists of two gauge symmetries; 4-dimensional diffeomorphisms and
internal gauge symmetry. We use light-cone gauge to fix a part of
redundancies. For that end, we use reparametrization of the
world-volume to identify the temporal direction of the brane with
the light-cone time $\tau \sim X^+$. Furthermore, in the low energy
limit the gauge field decouples but to make sure supersymmetry at
the quantum level in the spectrum of fluctuations, we need to give a
appropriate prescription for quantization. In the light-cone frame
the action reads as%
\be {\cal S}_{D3} = \int dX^+ L. \nn\ee%
The light-cone Hamiltonian is defined as ${\bf H} = P^- = -\partial
L/\partial\partial_\tau X^+$ which can be written
explicitly\footnote{Explicit derivation is given in
\cite{Hedge-Hogs,TGMT}.}%
\be\label{continuous-Hamiltonian}
\begin{split}
{\bf H} = \int{\rm d}^3\sigma&\bigg[\frac{1}{2p^+}(P_i^2+P_a^2) +
\frac{1}{2}\mu^2p^+(X_i^2+X_a^2) +
\frac{1}{2p^+((2\pi)^3l_s^4g_s)^2} \det g_{rs} \cr &-
\frac{\mu}{3!(2\pi)^3l_s^4g_s}\big(\epsilon^{ijkl}X^i\{X^j,X^k,X^l\}
+ \epsilon^{abcd}X^a\{X^b,X^c,X^d\}\big) \cr &+\frac{1}{2g_s}\left(
\theta^\dagger {}^{\alpha \beta} (\sigma^{ij})_\alpha^{\:  \:
\delta} \{X^i, X^j, \theta_{\delta \beta}\} + \theta^\dagger
{}^{\alpha \beta} (\sigma^{ab})_\alpha^{ \: \: \delta} \: \{X^a,
X^b, \theta_{\delta \beta}\}\right) \cr &+ \frac{1}{2g_s}
\left(\theta^\dagger {}^{\dot\alpha \dot\beta}
(\sigma^{ij})_{\dot\alpha}^{ \: \: \dot\delta} \: \{X^i, X^j,
\theta_{\dot\delta \dot\beta}\} + \theta^\dagger {}^{\dot\alpha
\dot\beta} (\sigma^{ab})_{\dot\alpha}^{\: \:
\dot\delta} \: \{X^a, X^b, \theta_{\dot\delta \dot\beta}\}\right) \bigg], \end{split}\ee%
where $p^+$ is the light-cone momentum which is defined as $p^+ =
-\partial L/\partial\partial_\tau X^-$, and $P^I$'s are transverse
momenta $P_I = \partial L/\partial\partial_\tau X^I$. We have
collected everything in the form of Nambu bracket%
\be \epsilon^{rst}\partial_r X^I\partial_r X^J\partial_r X^K \equiv
\{X^I,X^J,X^K\}, \ee%
and the determinant can be expanded as%
\be\begin{split} \det g_{rs} =
&\frac{1}{3!}(\{X^i,X^j,X^k\}\{X^i,X^j,X^k\} +
\{X^a,X^b,X^c\}\{X^a,X^b,X^c\}) \cr +&
\frac{1}{2}(\{X^i,X^j,X^a\}\{X^i,X^j,X^a\} +
\{X^a,X^b,X^i\}\{X^a,X^b,X^i\}). \end{split}\ee%

Another comment is in order. As can be seen from the light-cone
Hamiltonian, there is a unphysical pole for $p^+=0$ and so have to
be excluded. It can simply be done just by quantizing this
continuous variable, via $X^-$ compactification as%
\be X^- \sim X^-+2\pi R_- \Rightarrow p^+ = \frac{J}{R_-}\ . \ee%

In \cite{TGMT} it has been proposed that time-independent
volume-preserving diffeomorphism on connected, orientable and
compact manifolds is isomorphic to unitary group
$U(\infty)$.\footnote{with $U(1)$ factor as a trivial
diffeomorphism.} In other words, Nambu bracket on Reimannian 3
dimensional surfaces forms the algebra $su(\infty)$. The whole idea
is that this gauge group can be truncated and approximated by
$SU(J)$. It is performed by truncating infinite-dimensional harmonic
expansion of all the dynamical fields (3-brane coordinates) and
setting them in the adjoint matrix representation. The precise
discretization (regularization) prescription made in \cite{TGMT} is
as follows; replace functions with matrices or operators,
integration over spatial coordinates with trace of matrices and
Nambu brackets with four-commutators,{\it i.e.}%
\be\label{prescription}\begin{split} X^I \leftrightarrow
X^I_{J\times J}, \theta &\leftrightarrow \theta_{J\times J}, P^I
\leftrightarrow P^I_{J\times J} \cr p^+\int d^3\sigma
&\leftrightarrow \frac{1}{R_-} {\rm Tr} \cr \{{\cal F},{\cal
F},{\cal F}\} &\leftrightarrow J[{\cal O},{\cal O},{\cal O},{\cal
L}_5]\ .\end{split}\ee%
Then the regularized Hamiltonian for discretized D3-brane reads%
\be\label{Hamiltonian-2}\begin{split} {\bf H} = R_-\ {\rm Tr}&\Big[
\frac{1}{2}(P_i^2+P_a^2) +
\frac{1}{2}\left(\frac{\mu}{R_-}\right)^2(X_i^2+X_a^2) \cr &+
\frac{1}{2\cdot 3!g_s^2} \big([ X^i , X^j , X^k, {\cal L}_5][ X^i ,
X^j , X^k, {\cal L}_5] + [ X^a , X^b , X^c, {\cal L}_5][ X^a , X^b ,
X^c, {\cal L}_5]\big) \cr &+ \frac{1}{2\cdot 2g_s^2} \big([ X^i ,
X^j , X^a, {\cal L}_5][ X^i , X^j , X^a, {\cal L}_5] + [ X^a , X^b ,
X^i, {\cal L}_5][ X^a , X^b , X^i, {\cal L}_5]\big) \cr &
-\frac{\mu}{3!R_- g_s}\big( \epsilon^{i j k l} X^i [X^j, X^k, X^l,
{\cal L}_5]+ \epsilon^{a b c d} X^a [ X^b, X^c, X^d , {\cal L}_5]
\big) \cr &+\frac{1}{2g_s}\left( \theta^\dagger {}^{\alpha \beta}
(\sigma^{ij})_\alpha^{\:  \: \delta} [ X^i, X^j, \theta_{\delta
\beta}, {\cal L}_5] + \theta^\dagger {}^{\alpha \beta}
(\sigma^{ab})_\alpha^{ \: \: \delta} \: [ X^a, X^b, \theta_{\delta
\beta}, {\cal L}_5]\right) \cr &+ \frac{1}{2g_s}
\left(\theta^\dagger {}^{\dot\alpha \dot\beta}
(\sigma^{ij})_{\dot\alpha}^{ \: \: \dot\delta} \: [ X^i, X^j,
\theta_{\dot\delta \dot\beta}, {\cal L}_5]+ \theta^\dagger
{}^{\dot\alpha \dot\beta} (\sigma^{ab})_{\dot\alpha}^{\: \:
\dot\delta} \: [ X^a, X^b, \theta_{\dot\delta \dot\beta}, {\cal
L}_5]\right)
\Big]\ .\end{split}\ee%

In the prescription \eqref{prescription}, it is proposed that in
order to quantize Nambu 3-bracket one should introduce by hand a fix
(non-dynamical) matrix ${\cal L}_5$ and convert it to a well-defined
4-commutator. ${\cal L}_5$ is closely related to chirality operator
of $SO(4)$ and make it possible to survives trace and by-part
integration properties \cite{TGMT}. Starting from a Nambu 4-bracket
we can non-trivially fix one of the functionals in such a way that
it would not contribute and Nambu 4-bracket effectively reduces to
3-bracket but still with four functionals\footnote{If one fixes it
to identity matrix {\bf 1}, 4-commutator vanishes as opposed the
case of 3-commutator. This is one of the reasons why we refer to the
former well- and latter ill-defined.}. In looking for a unique and
basis independent definition of ${\cal L}_5$, we can always
diagonalize ${\cal L}_5$ using $U(J)$ rotation. Then Hermiticity and
tracelessness together with the fact that it squares to identity
matrix, fix the form of ${\cal L}_5$ to be a matrix with +1 and -1
eigenvalues up to permutation ${\cal S}_J$. It is worth noting that
although ${\cal L}_5$ is a non-dynamical matrix, it can take various
form or values \cite{TGMT,half-BPS}.

The Hamiltonian \eqref{prescription} of the 0+1-dimensional gauge
theory is written in the temporal gauge ${\cal A}_0 = 0$. To ensure
$SU(J)$ invariance, all the physical configurations must satisfy
${\cal A}_0$ equation of motion, which are $J^2-1$ independent
Gauss' law conditions.

Thus, the DLCQ of a D3-brane on the plane-wave background with a
compact null direction leads to a 0+1-dimensional supersymmetric
$U(J)$ gauge theory, a Matrix theory. It is conjectured to be the
DLCQ of type IIB string theory in the sector with $J$ units of
light-cone momentum on the AdS/plane-wave backgrounds \cite{TGMT}.
This theory, called tiny graviton Matrix theory (TGMT), is proposed
to be theory of $J$ tiny gravitons, the giant gravitons carrying
minimum light-cone momentum, which show brane structure and gauge
enhancement when some number of them sit on top of each other.

\subsection{Comparing the two proposals}
Thus far, we have presented two recipe to construct the DLCQ
Hamiltonian ${\bf H}_J^{\rm DLCQ}$ of type IIB string theory in the
sector $J$ on the AdS/plane-wave backgrounds. Both recipes result in
an unique Hamiltonian; the Hamiltonian coming from non-BPS D0-branes
\eqref{Hamiltonian-1} and the one coming from BPS D3-brane
\eqref{Hamiltonian-2} have exactly the same form. In particular, the
matrix ${\cal L}_5$ in the TGMT plays the same role as ${\cal T}$ in
our construction.

The similarity between these two different approaches for the
construction of the {\bf H}, shed light on the quantization
(regularization) of volume-preserving diffeomorphism on
3-dimensional Reimannian surfaces. Specifically, it gives an strong
evidence in favor of the discretization prescription introduced in
\cite{TGMT} and reviewed above, as the way the gauge group of
diffeomorphism is approximated by the unitary group $U(J)$.

\subsubsection{The symmetry structure}
The Matrix theory has a large number of local and global symmetries.
Although working in the temporal gauge, the Hamiltonian still enjoys
the time-independent part of the $U(J)$ gauge symmetry, which
appears as a global symmetry. Moreover, it has $PSU(2|2)\times
PSU(2|2)\times U(1)_{\bf H}\times U(1)_{p^+}$ supersymmetry group,
which is the supergroup of the plane-wave background, with the
minimal generators $Q_{\alpha\dot\beta},\ Q_{\dot\alpha\beta},\
J^{ij},\ J^{ab},\ {\bf H},\ p^+$. The Hamiltonian is invariant under
expected dynamical supersymmetries. The superalgebra is a big one
with 16 supercharges and 30 isometries. It puts severe restriction
on the form of Hamiltonian. A supersymmetric quantum mechanics with
16 supercharges is uniquely determined once we specify superalgebra
and the gauge group. Furthermore, its superalgebra naturally
contains some of the extensions. In particular it has a 4-form
corresponding to the dipole moment of the self dual RR 5-form,
needed for stabilization of the vacuum. For the supersymmetry
algebra and its matrix realized generators see \cite{extensions}.

The bosonic part of which is $SO(4)_i\times SO(4)_a\times U(1)_{\bf
H}\times U(1)_{p^+}\times {\mathbb Z}_2\times {\mathbb Z}_2$. The
two $SO(4)_i$ and $SO(4)_a$ rotations act on $i$ and $a$ vector
indices of the bosonic $X^i$ and $X^a$ fields and on the
spinor (Weyl) indices of fermionic $\theta_{\alpha\beta}$ as%
\be\label{SO4-rotation}\begin{split}
X^i_{rs}&\to \tilde X^i_{rs}= R^i_j X^j_{rs}\\
(\theta_{\alpha\beta})_{rs}&\to
(\tilde\theta_{\alpha\beta})_{rs}=R_{\alpha\gamma} (\theta_{\gamma\beta})_{rs}\\
{\cal T}&\to {\cal T} , \end{split}\ee%
where $R_{ij}=e^{i\omega_{ij} \gamma^{ij}}, \
R_{\alpha\gamma}=e^{i\omega_{ij} \sigma^{ij}}$ are respectively
$4\times 4$ and $2\times 2$ $SO(4)$ rotation matrices and $r,s$ are
$J\times J$ indices. $i$ and $a$ transverse directions are exchanged
by a ${\mathbb Z}_2$ symmetry. There is another $\mathbb{Z}_2$
symmetry which changes the orientation of the $X^i$ and $X^a$
simultaneously ({\it i.e.} $\epsilon_{ijkl},\ \epsilon_{abcd}\to
-\epsilon_{ijkl},\ -\epsilon_{abcd})$ together with sending ${\cal
T}\to -{\cal T}$.

Under the $U(J)$ rotations all the dynamical fields as well as the
${\cal T}$ are in the
adjoint representation:%
\be\label{U(J)-rotation}%
X_I\to U X^I U^{-1}\quad;\quad {\cal T}\to U {\cal T} U^{-1},\ee%
where $U\in U(J)$. There is a $U(1)$ subgroup of $U(J)$,
$U(1)_\alpha$ which is generated by ${\cal T}$:%
\be\label{Ualpha} U_\alpha = e^{i\alpha{\cal T}}. \ee%
This subgroup has the interesting property that keeps the ${\cal T}$
invariant.

\subsection{Evidence for the proposal}
Here we give some pieces of evidence in support of our Matrix theory
proposal. Having shown similarities with TGMT, it is a natural
expectation that they share in evidence.
\subsubsection{The vacuum structure}
The Hamiltonian relevant for the zero energy solutions, which are
static and bosonic, takes the form \eqref{positive-definite}. Each
term in that expression is positive-definite, hence the zero energy
solutions are obtained when each of the four terms are vanishing,
{\it i.e.}
\begin{subequations}\label{master-equations}\begin{align}
[X^i,X^j,X^k,{\cal T}] & = -\frac{\mu g_s}{R_-}\ \epsilon^{ijkl} X^l \\
[X^a,X^b,X^c,{\cal T}] & = -\frac{\mu g_s}{R_-}\ \epsilon^{abcd} X^d \\
[X^a,X^b,X^i,{\cal T}] &= [X^a, X^i, X^j,{\cal T}] =0\ .
\end{align}\end{subequations}

The first class of solutions to the above equations is the
trivial $X=0$ solution:%
\be\label{trivial} X^i = 0 \quad;\quad X^a = 0\ee%
Although mathematically trivial, as we will see this vacuum is
physically quite non-trivial.

The next class of solutions is obtained when either $X^i=0$ or
$X^a=0$. In this case equations.(\ref{master-equations}c) and either
of (\ref{master-equations}a) or (\ref{master-equations}b) are
trivially satisfied. Since there is a $Z_2$ symmetry in the exchange
of $X^i$ and $X^a$, here we only focus on the $X^a=0$ case and the
$X^i=0$ solutions have essentially the same structure. Therefore,
this class of vacua are solutions to
\be\label{single}%
X^a = 0 \ ,\ [X^i,X^j,X^k,{\cal T}] = -\frac{\mu g_s}{R_-}\
\epsilon^{ijkl} X^l\ .\ee%
In \cite{half-BPS} we gave the most general solutions to
\eqref{single}. One should, however, note that if we choose to
expand the theory around either of these vacua the $Z_2$ symmetry is
spontaneously broken. As we will see these solutions are generically
of the form of concentric fuzzy three spheres in either of the
$SO(4)$'s.

There is yet another class of solutions where both $X^i$ and $X^a$
are non-zero. These are non-trivial solutions which in the string
theory limit correspond to giant gravitons grown in both $X^a$ and
$X^i$ directions.

Noting the supersymmetry algebra of this Matrix theory
\cite{extensions}, it can be shown that these zero energy solutions,
either trivial or non-trivial, are half-BPS, {\it i.e.} they
preserve all of the dynamical supercharges (half out of whole
kinematical and dynamical ones), as they have ${\bf H}=0$ and ${\bf
J}_{ij}={\bf J}_{ab}=0 $ \cite{half-BPS}.

\subsubsection{Spectrum of fluctuations about the vacua}
The next evidence for our proposal comes from study of fluctuations
about above vacua. Fluctuations about the simplest vacuum, the
single fuzzy 3-sphere solution, has been worked out \cite{TGMT}. It
has been shown that it exactly matches that of a spherical D3-brane
giant graviton in the plane-wave background \cite{Hedge-Hogs}. The
effective coupling of
these fluctuation modes has also been worked. It is%
\be g_{eff} = \frac{R_-}{J\mu \sqrt{g_s}} =
\frac{1}{p^+\mu\sqrt{g_s}}\ee%
It is again what we obtain for spherical D3-brane giant graviton.

The fluctuations about the trivial vacuum has also been worked out
\cite{TGMT}. It has been shown that the spectrum of small BPS
fluctuations exactly matches with the spectrum of IIb supergravity
modes, BPS states of strings, on the plane-wave background
\cite{Metsaev}. The coupling of these fluctuation is%
\be g_{eff} = \frac{J^3}{(\mu p^+)^2g_s},\ee%
Based on this evidence, it has been conjectured that fundamental
type IIB strings are just non-perturbative objects about this vacuum
\cite{TGMT}.

\subsubsection{Analysis of BPS solutions} All the BPS states of this
Matrix theory have been studied and classified. Half-BPS states are
just vacua of the theory and are in the form of various
configuration of fuzzy 3-spheres or trivial ones \cite{half-BPS}.
Less-BPS stated have also been analyzed in details \cite{less-BPS}.
It has been shown that there is a one-to-one correspondence be-
tween these BPS states and those of $D = 4,{\cal N} = 4,U(N)$ SYM
theory \cite{Ferrara} and bubbling AdS geometries of IIb
supergravity \cite{LLM}.

\subsection{String theory limit} Physical
applications requires that the limit $R_-,J\rightarrow\infty$ be
taken while keeping fixed physical momentum $p^+$. It is called
string theory or decompactification limit. In this limit fuzzy
3-sphere generically goes over to round 3-sphere or spherical
3-brane giant graviton of radius%
\be R^2_{giant}/\alpha' = \left(\frac{g_s}{N}\right)^{1/2} J, \ee%
This relation between radius and angular momentum comes from
requirement of stability of topologically spherical BPS 3-brane. The
radius of fuzzy sphere comes from stable state of non-BPS D0-branes.
Together with%
\be R_-/\mu = (g_sN)^{1/2}\alpha',\ee these two relations coincide.
In this picture a BPS spherical D3-brane of radius $R$ is the string
theory limit of fuzzy 3-sphere, a state into which $J$ non-BPS
D0-branes are blown up.

Based on the study of fluctuation modes about the trivial vacua, it
has been conjectured that in the string theory limit this vacuum
quantum mechanically becomes the vacuum for strings on the
plane-wave background \cite{TGMT}. In this limit the coupling of
fluctuations become large and the fundamental type IIB closed
strings appear as non-perturbative objects in this vacuum.

It is interesting noting that the plane-wave background has two
$SO(4)$ isometries. D0-branes are singlets of $SO(4)$ which
eventually decay to other representations of this group which could
be either spherical or trivial(singlet) configurations. In fact,
this system stabilizes to either the configurations of fuzzy
3-spheres or trivial ones which in the string theory limit, go over
to giant gravitons or fundamental strings respectively.

\section{Conclusion and Outlook}\label{Conclusion}
Motivated by the tempting idea of discrete light-cone quantization
(DLCQ), in this note we proposed how to construct the DLCQ
Hamiltonian of type IIB string theory on the AdS/planewave
backgrounds in the sector with $J$ units of light-cone momentum. We
conjectured it is just the Hamiltonian of a 0+1-dimensional
supersymmetric $U(J)$ gauge theory, a Matrix theory. It is itself
the light-cone Hamiltonian for the stabilized phase of a system of
$J$ coincident non-BPS D0-branes of type IIB string theory on the
plane-wave background with a null circle. We presented some evidence
in support of this proposal.

Furthermore, on the one hand through the strong form of the
Maldacena's conjecture, quantum type IIB string theory on the AdS
background is equivalently described by a superconformal gauge
theory \cite{Maldacena,MAGOO}. From practical point of view, our
computational ability does not go beyond classical supergravity on
the AdS background.\footnote{For quantum string theory on the
plane-wave background we can go beyond this limitation \cite{BMN}.
However the question of quantum string theory on AdS background is
still there, which was the subject of this note.} On the other hand,
in this not we propose a Matrix theory which governs the quantum
dynamics of the very string theory by giving its DLCQ Hamiltonian.
It is now interesting to investigate the correspondence between
quantum string theory (quantum gravity) and gauge theory. Parts of
this analysis has been done \cite{half-BPS,less-BPS} and lots
remains to be done to complete this dictionary. It would be called
under the rubric MT-AdS/CFT correspondence.

The collection of unstable non-BPS D0-branes in the presence of RR
flux, stabilize via blowing up to a fuzzy 3-sphere, going over to
spherical D3-branes or giant gravitons in the string theory
(continuum) limit. The net RR charge of a spherical D3-brane is zero
but there is a non-zero electric or magnetic dipole moment of the
4-form field developed by the vacuum expectation value of the
tachyon, which stabilizes it against its tension. In fact, the
vacuum structures of this Matrix theory are configurations of fuzzy
3-spheres.

Recall that type IIB is related to M-theory through $T^2$
compactification. From eleven dimensional viewpoint, D3-branes are
just M5-branes wrapped on $T^2$. Following the same logic, we
propose that D0-branes are just M2-branes compactified on $T^2$, in
such a way that the resulting state is a unstable brane
\cite{progress}. On the other hand, a class of half-BPS solutions to
eleven dimensional supergravity has been constructed in
\cite{Bena-Warner} which correspond to M2-branes dielectrically
polarized into M5-branes and asymptotes to eleven dimensional $AdS$
space\footnote{See also \cite{LLM}, where solutions with regular
boundary conditions were found.}. We propose that this system is
related to ours via $T^2$ compactification when one of the circles
of the torus is interpreted as the tachyon field. In this sense
tachyon can be thought of as one extra spatial direction. It has
also been argued that fuzzy 3-sphere topologically is $S^3$ times
two points. The extra factor, controlled by ${\cal T}$, is
reminiscent of the 11th circle. The eleventh dimension of M-theory
which is hidden in perturbative string theory, manifests itself in
non-perturbative string theory through tachyon. This issue is still
under further study \cite{progress}.

Furthermore, M2-branes can also be compactified in a proper way to
lead to stable states, BPS KK gravitons. Hence, it is a natural
expectation that fuzzy 3-sphere, as regularized spherical D3-brane
giant graviton, could also be seen as bound state of point-like
gravitons of type IIB theory. Furthermore, as a theory of quantum
gravity, we suppose that the same Matrix theory could be derived
from a system of $J$ number of BPS gravitons or gravitational waves
of type IIB theory on the plane-wave background. In a paper in
preparation, we show that this configuration stabilizes again to a
fuzzy 3-sphere upon graviton condensation. In fact a microscopic
description of giant gravitons has been given in \cite{Lozano} but
it does not have the complete symmetry structure $SO(4)$.

In the sense of brane construction, one of the manifestation of the
brane democracy \cite{Townsend} is the ascending and descending
relation between D-branes. By justifying the existence of non-BPS
branes besides BPS ones and treating them on equal footing, we may
extend this democracy to encompass non-BPS branes. Consider general
form of the WZ part of the low energy effective action%
\be {\cal S}_{WZ}\sim\int_{M_{p+1}}{\rm Tr}{\cal P}\
e^{\iota_X\iota_X}C_n\big(\overleftarrow{\iota}_{[X,T]}+\wedge
DT\big)\wedge e^{DA+B}\nn.\ee%
It manifests coupling of gauge field $A$ and tachyon field $T$
living on and scalars $X$ transverse to the brane, to RR fields of
the bulk. Lower dimensional branes can be described in terms of
degrees of freedom of higher dimensional ones through $DA$ and/or
$DT$ and, conversely higher dimensional branes can be described by a
system of lower dimensional one in terms of $[X,T]$ and/or $[X,X]$.
Hence, D-branes are dynamically related by the process of open
string tachyon condensation and/or transverse fluctuations. In the
sense of ascending/descending relations among the branes, $A$ and
$X$ fluctuations relate branes with even steps and $T$ fluctuations
does with odd step. One would say that these processes lead to BPS
branes, but on the other hand non-BPS ones can be realized as decay
products in the process of annihilation of BPS branes and anti
branes \cite{non-BPS}. In this note starting from the lowest
dimensional branes\footnote{We would like to avoid Wick rotation to
D-instantons, we prefer Minkowskian space rather than Euclidean.} we
have constructed spherical D3-brane\footnote{Construction of
D$p$-branes from non-BPS D0-branes in flat background has been
studied in the earlier works \cite{Kluson2,AST}.}.

It is conceivable that both the BPS and non-BPS D$p$-branes of type
II theories have the same eleven dimensional origin and can be
treated on equal footing, {\it i.e.} they are BPS graviton, M2 or M5
branes which do or do not wrap around the extra spatial dimension
seen as tachyon. Hence, the existence of non-BPS D$p$-branes
suggests the presence of at least one more dimension in critical
string theory which is invisible in perturbation theory. In non-BPS
D$p$-branes tachyon looks as an extra transverse dimension to the
brane \cite{Kutasov}.

Hilbert space of Matrix quantum mechanics naturally contains multi
particle states. We may consider block-diagonal set of matrices
which obey classically independent equations of motion, as
describing two matrix theory objects \cite{Taylor}. This implies how
matrix theory encodes a configuration of multiple objects. In this
sense we can think of any matrix theory as a second-quantized theory
from the point of view of the target space. Furthermore, since this
Matrix theory is describing D3-branes, it is describing part of the
moduli which is non-perturbative with respect to fundamental string
theory \cite{TGMT}. Hence, this Matrix theory provides us with a
second-quantized non-perturbative formulation of type IIB string
theory.

As an equivalent description of a theory of quantum gravity, this
Matrix theory should also provide a realization of the holographic
principle. The plane-wave background has a 1-dimensional light-like
boundary. This 0+1-dimensional gauge theory can be a reliable
candidate for the holographic theory living on the boundary. It has
been proposed that Bekenstein's bound in terms of the DLCQ of
M-theory on backgrounds with a null Killing direction
with light-cone momentum $p^+$ reads \cite{Bousso}%
\be S_{DLCQ} \leq 2\pi^2 p^+ . \nn\ee%
It is interesting to test this proposal for the DLCQ of type IIB
theory on the ten-dimensional plane-wave background. Closely
related, it is also fascinating to study this Matrix theory at
finite temperature due to its relation to Hagedorn behavior and
black hole studies.

In the BFSS Matrix model for M-theory on the flat background, space
arises as moduli space of vacua of scalar field in the
0+1-dimensional D0-brane gauge theory \cite{Seiberg}. Vacuum
expectation values of which correspond to transverse coordinates of
the D0-brane. In this sense space is an emergent concept from more
fundamental degrees of freedom. We would like to extend this idea to
the case of Matrix model for type IIB theory on the AdS background,
where space arises from a 0+1-dimensional massive gauge theory.
Transverse coordinates arise properly, longitudinal light-cone
coordinate arises as the size of the matrices and as already
proposed, tachyon arises as eleventh dimension. Furthermore, it also
provides us with a suitable arena to study time-dependent phenomena
and may shed light on the concept and role of time.

In the context of AdS/CFT correspondence it has been shown that
non-BPS D0-branes are mapped to sphaleron solutions in the dual
gauge theory \cite{sphaleron}. They are unstable solutions located
at the saddle point of the potential in the configuration space, at
the top of a non-contractible loop. Here we showed that they
stabilize to spherical D3-brane giant gravitons. It would be
interesting to look for parallel process of tachyon condensation in
the gauge theory side.

For the type IIB string theory on the flat space, there is a
proposal for the Matrix formulation called IKKT Matrix model
\cite{IKKT}. It is based on a system of D-instantons and derived as
discretized Green-Schwarz action for fundamental string in the
Schild gauge . In order to relate our Matrix theory to IKKT model we
propose that in the absence of RR flux, non-BPS D0-branes decay and
leave out D-instantons as decay products. In the presence of the
flux, instead, they stabilized to another vacua which are fuzzy
3-spheres or discretized spherical D3-branes.

Finally, the most immediate and indispensable step to be taken in
favor of our Matrix theory conjecture, is to show that non-BPS
D0-brane is the only remaining and dominating degree of freedom for
the DLCQ of type IIB theory on the AdS space. This parallels
Seiberg-Sen argument for the DLCQ of M-theory on the flat background
\cite{SS}. We would like to generalize it to the non-flat (highly
curved) eleven and ten dimensional AdS backgrounds. It may also shed
light on the relation between corresponding plane-wave limit and the
DLCQ description. Parts of this argument is given in
\cite{Shomer,extensions} and appendix \ref{Appendix-AdS}, where it
is argued that how the DLCQ of a theory reduces its underlying
geometric and algebraic structure to corresponding Penrose limit and
Inon\"{u}-Wigner contraction, respectively. Further evidence in
support of this conjecture will be given in a future work
\cite{progress}.
\section*{Acknowledgments}
I am thankful to Shahin Sheikh-Jabbari for insightful discussions
and useful comments throughout the work. This work is supported in
part by the grant from Freydoon Mansouri Foundation.
\appendix
\section{Fuzzy 3-Sphere}\label{appendix-fuzzy-sphere}
In this appendix we review fuzzy 3-sphere and the way of
construction. More details can be found in \cite{half-BPS}.

Precise definition of round d-sphere is given in terms of
constraints on embedding coordinates containing two $SO(d+1)$
invariant tensors $\delta_{mn},\ \epsilon^{m_1m_2\dots m_{d+1}}$, as%
\bse\label{round-sphere}\begin{align} \sum_{m=1}^{d+1}\delta_{mn}X^mX^n &= R_d^2 \\
\{X^{m_1},X^{m_2},\dots,X^{m_d}\}_{N.B.} &=
R_d^{d-1}\epsilon^{m_1m_2\dots m_{d+1}}X^{m_{d+1}}\ . \end{align}\ese%
Then fuzzy d-sphere, for even $d$, is directly defined by
substituting continuous embedding functions with operator or
matrices and Nambu bracket with commutator, as
\bse\label{fuzzy-sphere}\begin{align} \sum_{m=1}^{d+1}\delta_{mn}X^mX^n &= R_d^2 \\
[X^{m_1},X^{m_2},\dots,X^{m_d}]_{D.B.} &=
i^{[d+1/2]}l^{d-1}J\epsilon^{m_1m_2\dots m_{d+1}}X^{m_{d+1}}\ .
\end{align}\ese%
For the case of odd-spheres, to overcome the difficulty of dealing
with odd-brackets, we prescribed to add one fixed matrix to convert
it into an even-bracket. For instance for fuzzy 3-sphere we have
\cite{TGMT,half-BPS}%
\bse\label{fuzzy3-sphere}\begin{align}
\sum_{i=1}^{4}\delta_{ij}X^iX^j &= R_3^2{\bf 1} \\
[X^i,X^j,X^k,{\cal L}_5] &= -l^2J\epsilon^{ijkl}X^l\ . \end{align}\ese%

The matrix equations \eqref{fuzzy3-sphere} may be solved by the
harmonic oscillator approach, according to which the embedding
coordinates of the fuzzy spheres are related to the coordinates of a
Moyal plane, $z_\alpha$ with harmonic non-commutative algebra
through Hopf map.%
\bea V^m_{\infty\times\infty} &=& \bar
z_\alpha(\gamma^m)_{\alpha\beta} z_\beta \\ N_{\infty\times\infty}
&=& \bar z_\alpha\delta_{\alpha\beta}z\ .
\eea%
where $\gamma$'s and $\delta$ are matrix Glebsch-Gordon
coefficients. First we would like to construct more handleable fuzzy
4-sphere by restricting to irreducible representation of $SO(5)$
which singles out $N\times N$ matrices inside infinite dimensional
matrices. It is done by the projection matrix ${\cal P}_N$ as%
\be W^m_{N\times N} = {\cal P}_N\bar
z_\alpha(\gamma^m)_{\alpha\beta} z_\beta{\cal P}_N\ . \ee%
One can easily check that the above coordinates truly satisfy
constraints of fuzzy 4-sphere \eqref{fuzzy-sphere}. Before moving to
the construction of fuzzy 3-sphere, we would like to comment on the
above construction of the 4-sphere. By definition an $S^4$ is a four
dimensional manifold with $so(5)$ isometries. In the above we have
given a specific embedding of a four sphere in an eight dimensional
(noncommutative) space. More specifically, noting that $N=const. $
defines an $S^7$ in the eight dimensional space, we have an
embedding of $S^4$ into $S^7$. This embedding is a (noncommutative)
realization of the Hopf fibration with $S^4$ as the base. Out of the
$so(8)$ isometries of the $S^7$ there is a $u(4)$ subgroup which is
compatible with the holomorphic structure on ${\mathbb C}^4\simeq
{\mathbb R}^8$. Note also that in the noncommutative Moyal case of
${\mathbb C}^4_\theta$, that is this $u(4)\subset so(8)$ which does
not change the noncommutative structure. The $X^\mu$ behaves as a
vector under $so(5)\subset su(4)$ and the generators of the full
$su(4)$ are $X^\mu$ and $[X^\mu,X^\nu]$. (The generator of the
$u(1)\subset u(4)$ is $N$).

Finally, we construct fuzzy 3-sphere via fixing one direction and
reducing one dimension. One should construct irreducible
representation of $SO(4)$ out of highly reducible one, by projecting
out to $J\times J$ matrices using the projector ${\cal P}_J$ as%
\bse\begin{align} X^i_{J\times J} &= {\cal P}_J{\cal P}_N\ \bar
z_\alpha(\gamma^i)_{\alpha\beta} z_\beta\ {\cal P}_N{\cal P}_J \\
{\cal L}_5 &= {\cal P}_J{\cal P}_N\ \bar
z_\alpha(\gamma^5)_{\alpha\beta} z_\beta\ {\cal P}_N{\cal P}_J \ .
\end{align}\ese%
one can easily check that above definition of coordinates satisfies
\eqref{fuzzy3-sphere}. These equations fully contains $SO(4)$
invariant matrices namely $\delta_{ij}, \epsilon^{ijkl}, {\cal
L}_5$.

Note that there is yet another equivalent solution to the matrix
equation. Suppose we take coordinates%
\be Y^i_{J\times J} = {\cal P}_J{\cal P}_N\ \bar
z_\alpha(i\gamma^5\gamma^i)_{\alpha\beta} z_\beta\ {\cal P}_N{\cal
P}_J \equiv i{\cal L}_5 X^i_{J\times J} \ee%
we can easily show that above definition of coordinates also
satisfies \eqref{fuzzy3-sphere}.%
\section{DLCQ on the AdS Space}\label{Appendix-AdS}
The DLCQ of a physical theory acts as Penrose limit and Inonu-Wigner
contraction on its underlying geometrical and algebraic structures
respectively \cite{Shomer,TGMT,extensions}. Indeed the parameter
$\mu$ characterizes a homotopy indicating a plane-wave setup, at
starting point of which ($\mu=0$)we have flat space.

One of the ingredients of the DLCQ is going to the frame of
light-cone (freely-falling) observer who gives the simplest possible
description. The other ingredient is the presence of a light-like
circle. The whole idea is that the compact null direction defined as
a limit of a space-like circle. In this appendix we see that how
these ideas apply on the AdS space and are encompassed in the
Penrose limiting process on the geometry. Parallel process at the
level of the algebra has been done in \cite{extensions}.

For that end consider $Z_M$ orbifolded $AdS_5\times S^5$ space in
the global coordinates with metric%
\be ds^2 = R^2\Big(-\cosh^2\rho d\tau^2+d\rho^2+\sinh^2\rho
d\Omega_3^2 + \cos^2\theta d\phi^2+d\theta^2+\sin^2\theta
d\tilde\Omega_3^2\Big) \nn\ee%
\be d\Omega_3^2 = d\alpha_1^2 +
\sin^2\alpha_1(d\alpha_2^2+\sin^2\alpha_2d\alpha_3^2) \qquad,\qquad
d\tilde\Omega_3^2 = \cos^2\psi d\chi^2 + d\psi^2 + \sin^2\psi
d\omega^2 \nn\ee%
\be \quad\ \chi\rightarrow\chi+\frac{2\pi}{M}\qquad,\qquad
\omega\rightarrow\omega-\frac{2\pi}{M}\ . \ee%
We focus on a null geodesic parameterized and defined by $\rho=0,\
\psi=0,\ \alpha=\frac{\pi}{2},\ \chi=\tau$, and define the
light-cone coordinates $x^\pm = 1/2(\tau\pm\chi)$. Orbifolding with
$Z_M$ on and boosting with rapidity $\beta$ along the two isometric
direction $\tau$ and $\chi$ changes light-cone coordinates as%
\be x^\pm\rightarrow e^{\mp\beta}\big(x^{\pm} + \pi/M \big), \ee%
then we rescale coordinate and take the limit $R\rightarrow\infty,\
\beta\rightarrow\infty,\ M\rightarrow\infty$ while keeping following
combinations fixed%
\be X^\pm\equiv Rx^\pm,\ x\equiv R\rho,\ y\equiv R\theta,\ z\equiv
R\psi,\ \frac{e^\beta}{R}\equiv\mu,\ \frac{e^\beta R}{M}\equiv 2R_-\
.\ee%
With this limiting process, Penrose limit, neighborhood of the
geodesic is stretched to become the whole space and the spacetime is
then reduced to its corresponding plane-wave background%
\be ds^2 = -2dX^+dX^- - \mu^2(x^2+y^2+z^2)(dX^+)^2 +
dx^2+x^2d\Omega_3^2 + dy^2y^2d\phi^2 + dz^2 + z^2d\omega^2, \ee%
it can be rewritten in a more convenient form as%
\be ds^2 = -2dX^+dX^- - \mu^2(X_i^2+X_a^2)(dX^+)^2 + dX_i^2 +
dX_a^2, \ee%
at the same time $X^-\rightarrow X^-+2\pi R_-$ compactifies and we
get a null circle as a limit of space-like circle. Similar limiting
process is applied to the form field. In this frame, spacetime is
parameterized by $X^+$, $X^-$ and $X^I$, the light-cone time,
light-cone longitudinal and transverse coordinates respectively. In
this basis momenta takes the form $p^+,P^-,P^I$ which are light-cone
momentum, light-cone Hamiltonian and transverse momenta.
%
\end{document}